\documentclass[conference]{IEEEtran}
\IEEEoverridecommandlockouts
\usepackage{cite}
\usepackage{amsmath,amssymb,amsfonts}
\usepackage{algorithm}
\usepackage{algorithmic}
\usepackage{todonotes}
\usepackage{graphicx}
\usepackage{textcomp}
\usepackage{xcolor}
\def\BibTeX{{\rm B\kern-.05em{\sc i\kern-.025em b}\kern-.08em
    T\kern-.1667em\lower.7ex\hbox{E}\kern-.125emX}}
\begin{document}

\title{Coordinated Autonomous Drones for Human-Centered Fire Evacuation in Partially Observable Urban Environments
}

\author{\IEEEauthorblockN{1\textsuperscript{st} Maria G. Mendoza} 

\IEEEauthorblockA{\textit{Mechanical Engineering} \\
\textit{University of California, Berkeley}\\
Berkeley, USA \\
maria\_mendoza@berkeley.edu}
\and
\IEEEauthorblockN{2\textsuperscript{nd} Addison Kalanther}
\IEEEauthorblockA{\textit{Electrical Engineering and} \\
\textit{Computer Sciences}\\
\textit{University of California, Berkeley}\\
Berkeley, California \\
addikala@berkeley.edu}
\and
\IEEEauthorblockN{3\textsuperscript{rd} Daniel Bostwick}
\IEEEauthorblockA{\textit{Electrical Engineering and} \\
\textit{Computer Sciences}\\
\textit{University of California, Berkeley}\\
Berkeley, USA \\
daniel.k.bostwick@berkeley.edu}
\and
\IEEEauthorblockN{4\textsuperscript{th} Emma Stephan}
\IEEEauthorblockA{\textit{Electrical Engineering and} \\
\textit{Computer Sciences}\\
\textit{University of California, Berkeley}\\
Berkeley, USA \\
estephan@berkeley.edu}
\and
\IEEEauthorblockN{5\textsuperscript{th} Chinmay Maheshwari}
\IEEEauthorblockA{\textit{Electrical and} \\
\textit{ Computer Engineering}\\
\textit{Johns Hopkins University}\\
Baltimore, USA \\
chinmay\_maheshwari@jhu.edu}
\and
\IEEEauthorblockN{6\textsuperscript{th} Shankar Sastry}
\IEEEauthorblockA{\textit{Electrical Engineering and} \\
\textit{Computer Sciences}\\
\textit{University of California, Berkeley}\\
Berkeley, USA \\
sastry@coe.berkeley.edu}
}

\maketitle


\begin{abstract}
Autonomous drone technology holds significant promise for enhancing search and rescue operations during evacuations by guiding humans toward safety and supporting broader emergency response efforts. However, their application in dynamic, real-time evacuation support remains limited. Existing models often overlook the psychological and emotional complexity of human behavior under extreme stress. In real-world fire scenarios, evacuees frequently deviate from designated safe routes due to panic and uncertainty.

To address these challenges, this paper presents a multi-agent coordination framework in which autonomous Unmanned Aerial Vehicles (UAVs) assist human evacuees in real-time by locating, intercepting, and guiding them to safety under uncertain conditions. We model the problem as a Partially Observable Markov Decision Process (POMDP), where two heterogeneous UAV agents—a high-level rescuer (HLR) and a low-level rescuer (LLR)—coordinate through shared observations and complementary capabilities. Human behavior is captured using an agent-based model grounded in empirical psychology, where panic dynamically affects decision-making and movement in response to environmental stimuli.

The environment features stochastic fire spread, unknown evacuee locations, and limited visibility, requiring UAVs to plan over long horizons to search for humans and adapt in real-time. Our framework employs the Proximal Policy Optimization (PPO) algorithm with recurrent policies to enable robust decision-making in partially observable settings. Simulation results demonstrate that the UAV team can rapidly locate and intercept evacuees, significantly reducing the time required for them to reach safety compared to scenarios without UAV assistance.\footnote{We provide access to more details for results and the simulation at \textbf{https://solid.eecs.berkeley.edu/humanitarian-response/}}

\end{abstract}
\vspace{0.5\baselineskip} 
\noindent
\begin{minipage}{\columnwidth}
\footnotesize
\raggedright
\textcopyright~2025 IEEE. Personal use permitted.
Accepted at \textit{IEEE Global Humanitarian Technology Conference (GHTC 2025)}.
Final version will appear in IEEE Xplore.
\end{minipage}


\section{Introduction}
Wildfires increasingly threaten urban areas, where dense infrastructure and limited escape routes make evacuation difficult. Recent events highlight how the lack of effective search and rescue systems has led to severe injuries and fatalities~\cite{Hoffman_2025}. Timely, informed response is critical, especially under high stress and low situational awareness. Incidents in densely populated regions, where people were trapped beyond safe evacuation windows due to limited rescue capacity~\cite{City_of_LA_Mayor_2025a}, underscore the urgency of integrating drone technology into disaster response—particularly in low-resource settings with constrained infrastructure and personnel.

Unmanned Aerial Vehicles (UAVs) and robotic systems are increasingly recognized as valuable assets in disaster response, offering affordability, mobility, and the ability to operate in hazardous conditions without endangering human lives~\cite{BCC_Research_2024}. While UAVs are already used for aerial mapping, environmental monitoring, and damage assessment, their role in active rescue and evacuation remains limited~\cite{Erdelj_HelpFromSky}. Yet, drones hold significant potential for more proactive use—such as locating and tracking individuals, providing real-time instructions, calming panic through interaction, and alerting rescuers to stranded victims.

Motivated by this, we study the following question:

\begin{quote}
How can a team of autonomous agents coordinate to search for and guide humans experiencing panic and stress during fire evacuation?
\end{quote}

We consider a dynamic urban environment where a fire spreads over time, focusing on a simplified scenario involving a single evacuee navigating to safety. This problem presents two main challenges:

\begin{itemize}
\item \textbf{Panic-induced behavior:} Panic and stress impair rational decision-making, leading to suboptimal or erratic actions~\cite{ALDAHLAWI2024systemreview, bakhshian2023evaluating, XENIDIS_BPDI}.
\item \textbf{Partial observability:} UAVs initially lack knowledge of the evacuee’s location, and urban occlusions further limit visibility.
\end{itemize}

To address these, we implement a coordinated strategy using two UAVs with asymmetric roles, extending the framework from~\cite{peg} to a humanitarian context:

\begin{itemize}
    \item The \textbf{High-Level Rescuer (HLR)} operates at a higher altitude to gain broad situational awareness and estimate the probable location of the human.
    \item The \textbf{Low-Level Rescuer (LLR)} navigates closer to the ground, allowing it to detect occluded regions, avoid obstacles, and interact with the evacuee to guide them to safety.
\end{itemize}

To model the evacuee, we adopt an agent-based model (ABM) approach as it offers a principled way to capture emergent behavior and heterogeneity in individual responses \cite{bakhshian2023evaluating, elhami2023human}. ABMs have been widely used in evacuation modeling to simulate how individuals deviate from optimal routes due to panic, limited visibility, or social influence. In this work, we use the panic model developed in~\cite{trivedi2018panic}.  

On the UAV side, we employ {}{deep reinforcement learning} to learn coordination strategies in a {}{partially observable} environment. Our framework trains a {}{single policy} to control asymmetric agents with different capabilities and objectives developed in \cite{peg}. To reason under uncertainty, we use {recurrent neural network-based policies} that leverage observation histories. We carefully design reward functions to balance exploration and exploitation during training, and use {}{Proximal Policy Optimization (PPO)} to train agents that can effectively assist humans during evacuation.

To evaluate our policies, we create testing environments with slight randomization in the start and end positions of the human evacuee, as well as the fire start locations. We observe that when the evacuees' start and end locations are the same or similar, the policy performs well regardless of the fire start locations. We also find that human panic plays a significant role in guiding evacuees out of dangerous situations. When a person panics or fails to notice the UAV that can lead them to safety, they significantly increase their time spent in hazardous zones.

\section{Related Works}
Aerial and ground robots have been increasingly deployed for humanitarian assistance and disaster response, yet their use in real-time search and evacuation operations remains limited~\cite{Erdelj_HelpFromSky}. Much of the existing literature focuses on post-disaster tasks such as aerial mapping and structural assessment~\cite{bendea2008low, italy2012rescue}, with relatively little attention to active human guidance during evacuations. One example of robot-assisted evacuation used tele-operated ground robots to guide individuals in smoke-filled indoor environments ~\cite{nayyar2023evacuee}. While this demonstrated the feasibility of robotic guidance, it relied on assumptions of proximity and prior knowledge of the human location and rational human behavior, limiting its applicability in more uncertain, outdoor, and dynamic environments.

Within robot-guided evacuation strategies, “shepherding"—where robots actively lead evacuees to exits—has been shown to outperform strategies where control is handed off between robots at key decision points~\cite{nayyar2019strategies}. However, such strategies require that robots first locate evacuees under uncertain conditions, including limited visibility and obstructed urban terrain. To address this, the literature on collaborative multi-robot systems offers several search techniques based on distributed exploration and information sharing~\cite{kruijff2015tradr, italy2012rescue, jorge2019survey, moosavi2024collaborative}. These methods can improve coverage and redundancy, thereby increasing the likelihood of locating distressed individuals in time. Yet, much of this work fails to fully integrate search with real-time guidance, especially in scenarios involving human unpredictability.

Finally, there is growing recognition that autonomous capabilities are crucial for robot deployment in disaster scenarios, particularly in low-resource settings where communication infrastructure may be unreliable or absent. Many systems still rely heavily on teleoperation~\cite{nayyar2023evacuee, kim2009portable}, which not only strains human operators but is also susceptible to failure under stress or scale~\cite{murphy2016disaster, nieto2014coordination}. Autonomous systems with onboard sensing, perception, and decision-making can offer more resilient and scalable solutions. However, autonomy must be paired with coordinated behavior across teams and with models that account for human panic and behavioral variability.

These gaps highlight the need for generalizable, autonomous, and coordinated multi-robot systems that can search for, reason about, and guide human evacuees in real-time. Our work builds on these insights by combining agent-based human modeling, deep reinforcement learning, and asymmetric drone roles to enable robust evacuation support in partially observable, dynamic environments.
\section{Problem Formulation}

We consider an urban environment affected by fire, involving three agents: a human agent (referred to as the \emph{evacuee}), who aims to reach a designated safe zone while avoiding the fire, and a team of UAVs (referred to as \emph{rescuers}), whose objective is to locate and guide the evacuee to safety. The rescuers operate collaboratively, with one responsible for high-level surveillance and the other for ground-level interception. We describe each rescuer below:

1. \textbf{High-Level Rescuer (HLR):} Operates at high altitude with a wide field of view. It maps the environment, infers the evacuee’s location, and tracks their movement using a downward-facing camera. However, it cannot physically intercept the evacuee.

2. \textbf{Low-Level Rescuer (LLR):} Operates at low altitude with a narrower field of view. It is the only agent capable of physically intercepting the evacuee. The LLR navigates closer to the ground and detects obstacles that may be hidden from the HLR, using a front-facing camera.

\begin{figure}[!h]
    \centering    \includegraphics[width=0.5\textwidth]{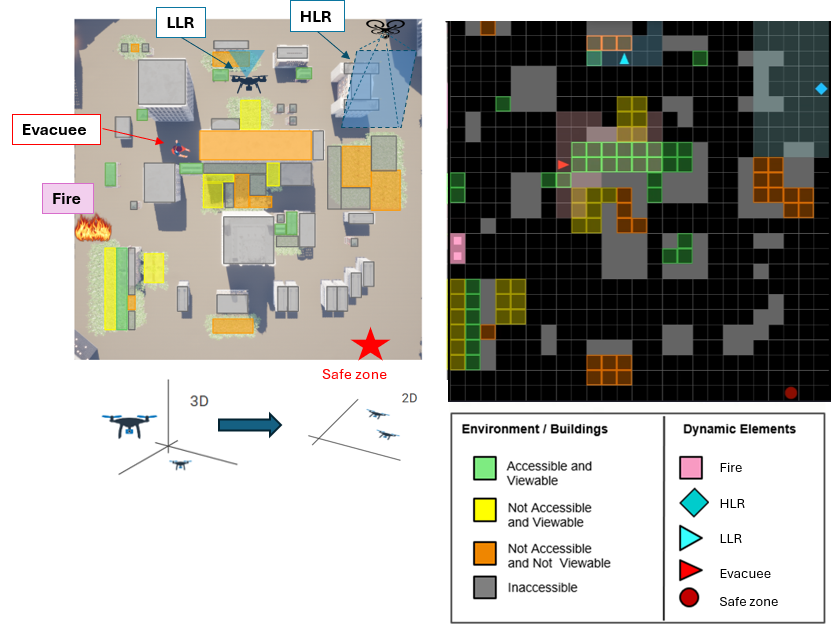}
    \caption{\textbf{Left:} A 3D urban environment from a top-down perspective illustrating a disaster evacuation scenario involving a human evacuee, fire, and two UAV rescuers. \textbf{Right:} The same environment is modeled as a 2D grid-based world, where each cell is annotated with accessibility and visibility properties. }
    \label{fig:urbanEnv}
\end{figure}

We model the urban environment as a discrete-time, 2D grid world, represented by a tuple $\mathcal{G} = (\mathcal{X}, \mathcal{A}, \mathcal{O}, \mathcal{B})$, as shown in Fig. \ref{fig:urbanEnv}. Our grid design extends the base environment introduced in~\cite{peg} by incorporating a dynamic fire model where:
\begin{itemize}
    \item $\mathcal{X} \subseteq \mathbb{Z}^2$ denotes the set of grid cells, divided into accessible and inaccessible regions.
    \item $\mathcal{A}$ defines the set of possible actions (e.g., move up / down / left / right, wait).
    \item $\mathcal{O}$ encodes the local observations available to each agent (e.g., visibility, fire, evacuee location).
    \item $\mathcal{B}_t \subset \mathcal{X}$ denotes the subset of cells affected by fire at time $t$.
\end{itemize}

Fire propagates over time using a simple stochastic model and can be adapted to other models\footnote{Even though we assume a simple fire spread model, our design approach is modular and can incorporate other complex fire models as well.}. At each time step $t$, each burning cell $(x, y) \in \mathcal{B}_t$ may ignite one of its adjacent neighbors $(x', y') \in \mathcal{N}(x, y)$, where $\mathcal{N}(x, y)$ denotes the neighborhood of the cell \((x,y)\). Each neighbor cell $(x', y')$ ignites independently with probability $p_{\text{fire}}$, provided that it is within the map limits and not already burning. The probability that a given cell ignites is defined as 

\begin{align}
&\mathbb{P}\big((x', y') \in \mathcal{B}_{t+1} \,\big|\, (x', y') \notin \mathcal{B}_t\big) = 
1 -\nonumber \\ &\quad \quad \prod_{(x, y) \in \mathcal{N}(x', y')} 
\left(1 - \mathbf{1}_{\{(x, y) \in \mathcal{B}_t\}} \cdot p_{\text{fire}}\right). 
\end{align}

Each cell may contain static obstacles (e.g., walls, buildings, trees) or dynamic entities (e.g., agents, fire, smoke). The environment transitions based on both agent actions and environmental dynamics. Static obstacles are classified into four types based on two agent-specific properties: accessibility\footnote{An obstacle is accessible if an agent can occupy its cell. The HLR is unaffected by accessibility, as it flies above all structures.} and viewability\footnote{An obstacle is viewable if an agent can detect another agent through it when within its FOV.}. These properties vary across the HLR, LLR, and the evacuee:

\begin{itemize}
  \item \textbf{Type I:} Inaccessible and non-viewable by all agents (e.g., solid walls or opaque buildings).
  \item \textbf{Type II:}  Inaccessible and non-viewable by the HLR, but accessible and viewable by the LLR and the Evacuee (e.g., tree rows, tall canopies, or partially open structures).
  \item \textbf{Type III:} Accessible and viewable by the Evacuee; viewable but inaccessible by the LLR (e.g., narrow alleys or dense urban corridors).
  \item \textbf{Type IV:} Accessible and viewable by the Evacuee, but both inaccessible and non-viewable by the LLR and HLR (e.g., enclosed indoor areas).
\end{itemize}

Obstacles of Types I–IV occlude an agent’s field of view when not viewable by that agent type. In Fig. \ref{fig:urbanEnv}, obstacle types include: Type I (gray, inaccessible), Type II (orange, not accessible and not viewable), Type III (yellow, not accessible but viewable), and Type IV (green, accessible and viewable). Fire sources (pink tiles) dynamically spread through the environment. The human evacuee (red triangle) must reach the designated safe zone (red circle in 2D, red star in 3D) while avoiding fire and inaccessible static obstacles (Type I). The HLR (blue diamond) and the LLR (blue triangle) must navigate the partially occluded environment to search and intercept the evacuee. Shaded regions around each UAV indicate their respective fields of view (FOV).

The HLR and LLR share their positions, headings, fire observations, and detected information about evacuees. The HLR communicates any sightings to the LLR to enable coordinated interception. The objective is to intercept the evacuee as quickly as possible. 

The evacuee follows a behavior model shaped by panic, uncertainty, and environmental risk, including uncertain responses such as hesitation or rerouting. This creates a dynamic and partially observable environment requiring adaptive strategies from the UAV agents. 

In this work, we address the following key challenges:

{{}\textbf{C1}}  \textbf{Human Behavior under Panic.}  
The rescuers must be able to search and intercept the evacuee, especially because evacuees can act irrationally under stress with actions influenced by panic, limited information, {and environmental signals.} 

{{}\textbf{C2}} \textbf{Long-Term Planning under Uncertainty}. UAVs must plan over long horizons despite occlusions and evolving conditions to gather information, track, and intercept the evacuee. This requires modeling a partially observable environment in which the rescuer only relies on their observations to plan their actions. 

{{}\textbf{C3}}  \textbf{Time-varying environment}. The fire evolves stochastically, and its behavior is unknown to all agents, influencing both agent behavior and human decision-making. UAVs must adapt to changing fire conditions with limited foresight.

{{}\textbf{C4}}  \textbf{Collaborative Multi-Agent Coordination}: rescuers must coordinate their actions and share partial observations to track and intercept the evacuee effectively. Previous work has shown that naive and uncoordinated deployments of UAVs often fail to guarantee successful search and rescue outcomes in complex environments \cite{Erdelj_HelpFromSky}. Coordinated planning and communication are essential to achieving critical time intervention in dynamic and uncertain settings.

{{}\textbf{C5}}  \textbf{Heterogeneous Agent Capabilities}. The HLR and LLR differ in observation and action modules. Coordinating asymmetric agents poses challenges in policy learning.

{{}\textbf{C6}} \textbf{Unknown evacuee location.} The initial position of the evacuee is unknown to both the HLR and LLR agents, requiring them to actively search the environment.

\section{Approach}\label{approach_sec}

In this section, we detail the modeling components of our framework and explain how we address challenges \textbf{{(C1)-(C6)}}.

\subsection{Agent-Based Modeling with Panic Behavior} \label{human_modeling}
To address {{}\textbf{C1}}, we implement an agent-based model where a single human agent (the evacuee) interacts with a dynamic environment and makes decisions influenced by cognitive and emotional factors. We incorporate panic as a key psychological variable that impacts rational decision, following their agent-based model of human evacuation under panic developed by Trivedi and Rao \cite{trivedi2018panic}. 

The evacuee follows a rule-based policy with a predefined goal, simulating awareness of a potential safe zone. 
Under normal conditions, it plans movement using a grid-based path. 
However, the panic parameter dynamically modulates behavior: elevated panic may override rational planning, leading to irrational responses such as freezing or erratic motion. 
For simplicity, we currently model social forces as an exogenous factor rather than through explicit agent-to-agent interactions.

The evacuee’s observations include its position, goal, fire visibility, and last known rescuer locations, which feed into both panic updates and path decisions. At each time step \( t \), the panic stimulus \( \gamma(t) \in [0,1] \) is computed as follows 
\[
\gamma(t) = \frac{1}{2} (\gamma(t-1) + \delta(t)), 
\]
where 
\[
\delta(t) = \frac{1}{4} \sum_{k=1}^4 \delta_k(t), \delta(t) \in [0,1]
\]
and
\begin{align*}
\delta_1(t) & = \text{distance-to-exit factor (0 if near exit, 1 if far)}; \\
\delta_2(t) & = \text{misalignment with neighbors' velocity}; \\
\delta_3(t) & = \text{presence of nearby fire};\\
\delta_4(t) & = \text{presence of nearby agents in discomfort}. 
\end{align*}
In our simulation, $\delta_1$ is computed based on Euclidean distance to a known safe zone. To account for natural variability in human responses, $\delta_2$ and $\delta_4$ are sampled from normal distributions. $\delta_2$ reflects uncertainty or inconsistency in interpreting surrounding motion (e.g., misalignment with others' behavior) and $\delta_4$ reflects hesitation or delayed reactions to stress-inducing visual cues (e.g., seeing others in discomfort). Lastly, $\delta_3$ is deterministically set when fire is within the evacuee’s field of view (FOV).

The evacuee’s velocity $v(t) \in \mathbb{Z}^2$ is then updated as follows:

\[
v(t) = (1 - \gamma(t)) \cdot v_{\text{optimal}} + \gamma(t) \cdot v_{\text{herd}}
\]

where \( v_{\text{optimal}} \) is the velocity directed towards the goal, while \( v_{\text{herd}} \) aligns with the average heading of the local neighbors. Finally, the position $p(t) \in  \mathbb{Z}^2$ is updated using:

\[
p(t+1) = p(t) + v(t) \cdot \Delta t
\]

\subsection{Partial Observable Markov Decision Process (POMDP)} 
To address {{}\textbf{C2-C3}}, we model the environment as a POMDP represented by the tuple
\[
\mathcal{M} = (\mathcal{S}, \mathcal{A}, \mathcal{O}, T, \mathcal{R}, \gamma),
\]
where $\mathcal{S}$ is the latent environment state (comprising the location of all agents, location of fire, field of view of all agents), including the true positions of all agents, fire spread, and the evacuee's internal panic level. This state is not directly accessible to the agents. $\mathcal{A} = \mathcal{A}^{\text{HLR}} \times \mathcal{A}^{\text{LLR}}$ is the joint action space and $\mathcal{O} = \mathcal{O}^{\text{HLR}} \times \mathcal{O}^{\text{LLR}}$ is the joint observation space, $T$ is the stochastic transition function capturing UAV and fire dynamics, $\mathcal{R}$ is the reward function, and $\gamma \in (0,1)$ is the discount factor.
Each agent $i \in \{\text{HLR}, \text{LLR}\}$ selects an action in polar form, which defines a target point relative to its current position. The environment then computes an $A^*$ path to its next waypoint and truncates it according to the agent's speed. The LLR's action space is defined as:
\[
a_t = (r, \theta, \psi), \quad r \in [0, R],\ \theta, \psi \in [-\pi, \pi],
\]
where $(r, \theta)$ defines the target point and direction in polar coordinates within selection radius $R$, and $\psi$ specifies the heading direction. 

\paragraph{Observation Space.} At time $t$, each rescuer receives a partial observation of the environment. Observations for the HLR and LLR are defined as:
\begin{align*}
o_t^{\text{HLR}} &= o_t^{\text{LLR}} = \Big\{ 
     \mathcal{F}_t^{\text{HLR}},\ 
    p_t^{\text{HLR}},\ 
    p_t^{\text{LLR}},\ 
    \phi_t^{\text{LLR}}, \\
    &\quad (p_t^{\text{evac}},\ \phi_t^{\text{evac}}) \cdot 
    \mathbf{1}\left[
        p_t^{\text{evac}} \in 
        \mathcal{F}_t^{\text{HLR}} \cup \mathcal{F}_t^{\text{LLR}}
    \right]
\Big\},
\end{align*}
where 
$\mathcal{F}_t^i \subseteq \mathcal{X}$ denotes the field of view (FOV) of agent $i \in \{\text{HLR}, \text{LLR} \}$ at time $t$, and $p_t^i, \phi_t^i$ are its position and orientation. Rescuers observe their own states, each other’s positions, LLR's heading, and the evacuee’s position $p^{\text{evac}}$ and orientation $\phi^{\text{evac}}$, when visible to either agent. Each agent maintains a history of past observations and actions
\[
h_t^i = \{(o_{t'}^i, a_{t'-1}^i) \mid t' \leq t\},
\] The team policy is condition on the joint history $h = (h_t^{HLR},h_t^{LLR})$
and acts according to a policy $\pi: h_t \mapsto a_t$, such that:
\[
a_t = \pi(h_t).
\]

Under partial observability, we train each policy $\pi_\theta$ using Proximal Policy Optimization (PPO) with recurrence, where $h_t$ is encoded via an LSTM. Each policy is trained to maximize the expected cumulative reward:
\[
J(\theta = \mathbb{E}_{\pi_\theta} \left[\sum_{t=0}^{T} \gamma_t \text{reward}_t \right].
\]

\subsubsection{Reward Design}
To guide the LLR and HLR agents toward the evacuee, we implement a reward function that promotes visibility, proximity, and successful capture. The rescuers share a reward and are trained jointly to encourage coordination. The reward is computed as:

\begin{align*}
    &\mathcal{R}_{\textsf{LLR/HLR}}(s, a, s') \\ &= 
    \begin{cases}
        c_{capture}, \  \text{if} ~||p_{\textsf{LLR}} - p_{\textsf{evac}}|| < r_{capture}  \\
        \alpha(l_0 c_{0}+ l_1 c_{1} + l_2 c_{2} + c_3 l_3), \ \text{otherwise}
    \end{cases}
\end{align*}

where:
\begin{itemize}

    \item $l_{0}= \mathbf{1}[\text{id}_{\text{evac}} \in \mathcal{F}^{\textsf{HLR}}], l_{1}= \mathbf{1}[\text{id}_{\text{evac}} \in \mathcal{F}^{\textsf{LLR}}]$:  indicator rewards for the evacuee being in the HLR or LLR’s FOV, respectively. 
    \item $l_{2}= \mathbf{1}[\text{evac seen in $s$ and $s'$}] \cdot (||\hat{p}_{\text{LLR}} - \hat{p}_{\text{evac}}||_2 - ||p_{\text{LLR}} - p_{\text{evac}}||_2)$: small reward for approaching the evacuee
    \item $-l_3$: current timestep $t$ in the episode
    \item $\alpha$: $\frac{1}{[\text{LLR Max Speed}]}$: scaling coefficient for speed
\end{itemize}

We set the following parameters $c_0 = 1$, $c_1 = 1$, $c_2 = 1$, $c_3 = 0.05$, and $c_{\text{capture}} = 10$ in our simulations to balance exploration and exploitation, and $r_{capture}$ as a very small number. Using a POMDP and the reward structure, we encourage the agents to explore the environment when the evacuee's location is unknown. This addresses {\textbf{C6}}, as the rescuers must rely on their partial observations to locate the evacuee. The HLR learns to maximize aerial visibility, while the LLR navigates through accessible regions to increase ground-level coverage. The designed reward incentivizes coordinated movement that balances exploration and pursuit.

\subsection{Centralized Multi-Agent Reinforcement Learning} 

To address {{}\textbf{C4-C5}}, we adopt a centralized training and execution framework in which the two rescuer agents share observations during both training and deployment. The team is controlled by a joint policy based on a recurrent neural network to leverage the history of observations, and is optimized using Proximal Policy Optimization (PPO). The choice of centralized training is motivated by the fact that the agents operate with asymmetric roles: the HLR is responsible for wide-area search and tracking, while the LLR, focuses on close-range interception. This heterogeneity introduces coordination and optimization challenges due to differing behavioral objectives. These factors make it difficult to attribute success or failure to individual actions, often leading to unstable policy updates and increased sample complexity during training. Thus, a joint policy utilizing each other's observations simplifies challenges \textbf{C4-C5} since each agent learns to coordinate with each other under the shared framework.

\section{Analysis}
We conducted simulations to investigate the following three research questions:
\begin{itemize}
\item[\textbf{Q1}] What is the effect of UAV support on an evacuee’s ability to reach the safe zone more quickly?
We assume that once a UAV reaches the evacuee, the evacuee follows the UAV to the destination via an \(A^\star\)-computed path (i.e., a rational path).
\item[\textbf{Q2}] How effective is the deep reinforcement learning based approach to learn a policy for UAVs?

\item[\textbf{Q3}] How robust is our framework to variations in the environment, the start locations of the evacuee, the safe zone, and the fire's origin?
\end{itemize}

To address these questions, we consider a training environment in which the fire can randomly originate in one of four distinct locations spread around the map symmetrically, and the evacuee start location and the safe zone is spread around the perimeter of the environment with 40 different start-end pairs, forcing the evacuee to traverse all regions of the map during training. This information is initially unknown to the rescue team. Further details about the algorithm and training parameters are provided in Appendix~\ref{appendix:training}.

\begin{figure}[!h]
    \centering
    \includegraphics[scale=0.32]{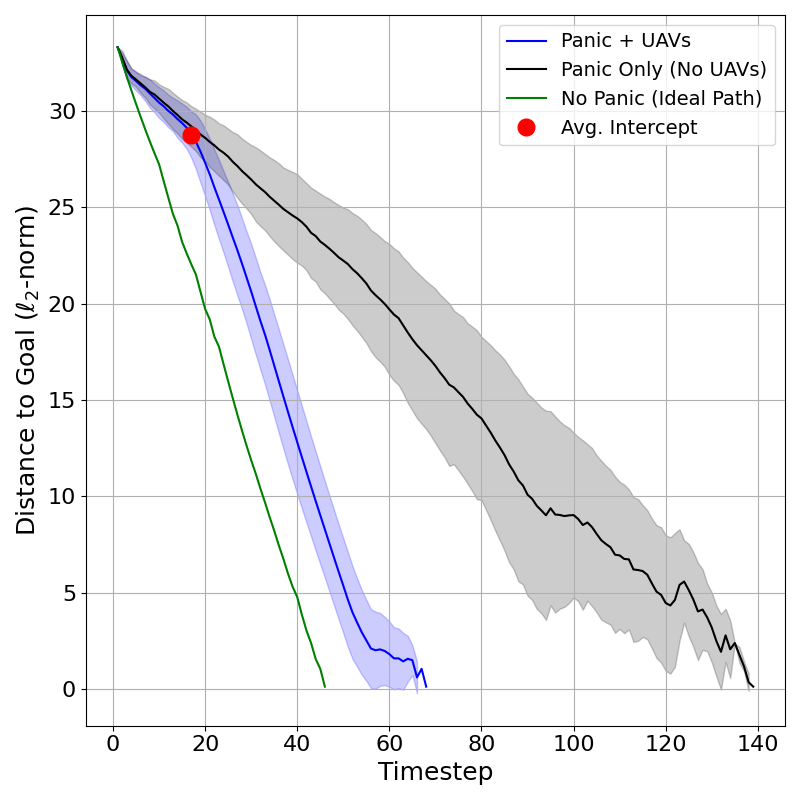}
    \caption{Number of steps taken by the evacuee to reach the safe zone under varying panic levels}
    \label{fig:ImprovementUAV}
\end{figure}

To address \textbf{Q1}, Figure \ref{fig:ImprovementUAV} presents results from 100 simulations. Following the parameters described in Section \ref{human_modeling}, we set the distributions for $\delta_2$ and $\delta_4$ as $\mathcal{N}(0.2, 0.1^2)$. These choices reflect evacuees who are, on average, mildly perturbed with panic contributions rarely exceeding 0.5 unless other factors like fire visibility, $\delta_3$ increases. The average evacuee trajectories are shown with shaded regions indicating variability in panic behavior. We compare three scenarios: (1) a baseline trajectory representing rational behavior without panic (green), (2) a trajectory where the evacuee remains in panic throughout (black), and (3) a trajectory where the evacuee starts in panic but is intercepted and guided by the LLR (blue), with the interception point marked by a red dot. On average, the rescuers take approximately 17.5 timesteps to locate and engage the evacuee. As expected, persistent panic nearly doubles the time required to reach the safe zone compared to the rational baseline, while UAV guidance significantly reduces evacuation time despite initial panic.

To study \textbf{Q2} and \textbf{Q3}, we increase the complexity of the environment by varying the initial conditions in four steps:
\begin{itemize}
    \item \textbf{Env-I}: Identical to the training environment. The evacuee’s start location, safe zone location, the origin of the fire, remain fixed.
    \item \textbf{Env-II}: The evacuee’s start location and safe zone location are randomly generated within a \textit{position randomization radius} $r$ of those in \textbf{Env-I}, while fire locations remain fixed as in \textbf{Env-I}.
    \item \textbf{Env-III}:  The initial conditions are the same as in \textbf{Env-I}, but fire origin is randomly generated within radius $r$ of those in \textbf{Env-I}.
    \item \textbf{Env-IV}: Both the evacuee’s start and safe zone locations, and the fire origin is randomly generated within radius $r$ of the corresponding positions in \textbf{Env-I}. 
\end{itemize}

We evaluate the following performance metrics to assess the robustness and effectiveness of our approach:
\begin{itemize}
\item \textbf{Rescuer Capture Rate:} The percentage of trajectories in which the LLR successfully captures the evacuee.
\item \textbf{Time to Capture:} The number of steps it takes for the rescuer team to intercept the evacuee.
\item \textbf{First Seen:} The amount of time it takes for the evacuee to be seen by the rescuer team for the first time within a scenario.
\item \textbf{Time in FOV:} The percentage of timesteps during which the evacuee is within the FOV of either rescuer. This reflects the rescuers' ability to track panic-influenced behavior. This metric also considers only trajectories where the evacuee is observed.
\end{itemize}

\begin{table*}[t]
\centering
\caption{Policy Performance Summary}
\small
\begin{tabular}{|c|p{2.5cm}|p{2.5cm}|p{2.5cm}|p{2.5cm}|}
\hline
\textbf{Env} & 
\textbf{Win Rate (\%) \(\uparrow\)} & 
\textbf{Capture Time \(\downarrow\)} & 
\textbf{First Seen \(\downarrow\)} & 
\textbf{FOV Time (\%) \(\uparrow\)} \\
\hline
\textbf{I (0, 0)} & 
\textbf{71.0 $\pm$ 4.54} & 
13.20 $\pm$ 1.17 & 
8.48 $\pm$ 0.97 & 
30.77 $\pm$ 1.66 \\
\hline
\textbf{II (5, 0)} & 
61.0 $\pm$ 4.88 & 
12.28 $\pm$ 0.59 & 
7.46 $\pm$ 0.35 & 
31.60 $\pm$ 1.70 \\
\hline
\textbf{III (0, 5)} & 
68.0 $\pm$ 4.66 & 
\textbf{10.91 $\pm$ 0.51} & 
\textbf{6.75 $\pm$ 0.30} & 
\textbf{33.77 $\pm$ 1.83} \\
\hline
\textbf{IV (5, 5)} & 
66.0 $\pm$ 4.74 & 
12.83 $\pm$ 1.17 & 
8.77 $\pm$ 1.14 & 
32.79 $\pm$ 1.88 \\
\hline
\end{tabular}
\label{tab:pursuer_evader_metrics}
\end{table*}

Table \ref{tab:pursuer_evader_metrics} shows that the rescuer team achieves its highest overall win rate in Env-I, when the origin of the evacuee, location of the safe zone, and fire origin remain unchanged from its training environment. Furthermore, the performance remains comparable across perturbed environments, demonstrating robustness to changes in initial conditions and addressing \textbf{Q3}.

Notably, in \textbf{Env-III}, the learned policy yields the best performance in terms of capture time (\textit{rescued time}), time to “first seen", and time spent in the rescuer's FOV. The location of the fire changes the trajectory of the evacuee, which leads to better performance in terms of these metrics. 

Furthermore, we define \textbf{Percent Improvement} as the ratio of the difference between the time it takes for the evacuee to reach the safe zone without rescuers and with rescuers, to the difference between the time it takes for the evacuee to reach the safe zone without rescuers and that of the optimal \(A^\star\) trajectory (evacuee with no panic). A higher value of this metric indicates that introducing the rescuer helps in evacuation.

\begin{figure}[!h]
    \centering
    \includegraphics[scale=0.38]{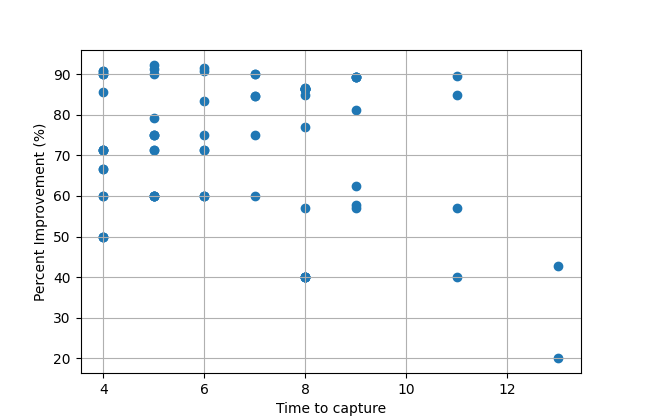}
    \caption{Time to capture vs. percent improvement in evacuation time}
    \label{fig:scatterplot}
\end{figure}

Figure~\ref{fig:scatterplot} shows the relationship between capture time and percent improvement when UAV rescuers are deployed. The evaluation is conducted in \textbf{Env-IV}, with randomized evacuee start positions, safe zone location, and fire origins within a position randomization radius of $r=5$. The plot includes only scenarios in which the evacuee was successfully captured, accounting for 77\% of all cases. We observe that, in most of these captured instances, the percent improvement exceeds \(60\%\), showing the effectiveness of the rescuers in aiding evacuation.

\begin{figure}[!h]
    \centering
    \includegraphics[scale=0.32]{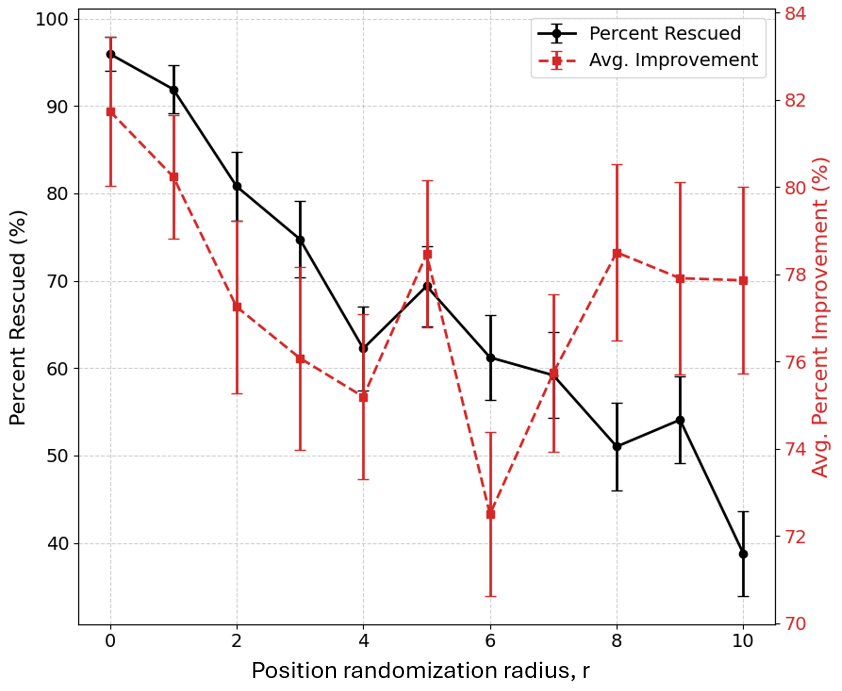}
    \caption{Impact of $r$ on evacuation outcomes. The left axis shows the percentage of evacuees successfully intercepted by UAV rescuers. The right axis shows the average percent improvement in capture time relative to the panic-only baseline.}
    \label{fig:perc_impr}
\end{figure}

Next, we study the robustness of our approach by varying the values of \(r\) in \textbf{Env-IV}. Figure~\ref{fig:perc_impr} compares two key metrics: (1) the percentage of episodes in which the rescuers successfully intercepted the evacuee, and (2) the average percent improvement, computed over successful rescues. We vary \(r\) from 0 to 10. When \(r = 10\), the evacuee can originate from any grid point and travel to any other grid point. As expected, the rescue rate decreases as \(r\) increases, since the trained policy encounters scenarios that deviate more from the training distribution. However, even when \(r = 10\), the evacuee is rescued in over \(40\%\) of the runs. Moreover, the average percent improvement remains above \(70\%\). These results suggest that the learned policy remains effective in diverse environments that deviate substantially from the training environment, highlighting the robustness of our approach. To further improve performance in such environments, one promising direction is to increase the diversity of origin-destination grid points and fire locations used during training.

\section{Conclusion}

We proposed a novel framework for autonomous fire evacuation support using a coordinated team of UAVs operating in a partially observable and dynamically evolving urban environment. Our approach addresses several key challenges in real-world evacuation scenarios, including uncertainty in the evacuee's location, limited observability due to occlusions, and the impact of panic on human decision-making. Simulation results demonstrate that UAV support significantly improves evacuation outcomes by reducing time to safety and compensating for suboptimal human responses. 

This technology is envisioned as a complementary tool for existing emergency response systems, with potential deployment by local fire departments or emergency agencies, particularly in settings where limited personnel and situational awareness constrain timely response.

A key limitation of our framework lies in several modeling assumptions, including perfect communication between UAVs, a simplified urban environment, and an abstracted representation of human behavior. However, to the best of our knowledge, this work is among the first to integrate a multi-UAV coordination strategy with a psychologically grounded human behavior model for humanitarian applications. While our panic model is grounded in empirical literature and captures key aspects of stress-induced decision-making, we recognize the broader complexity of human behavior in emergency contexts. Future work could explore alternative modeling approaches, including rule-based social dynamics or data-driven behavioral models.

Other promising directions are scaling to multi-human scenarios with $N$ evacuees and $M$ rescuers. While the current framework assumes $M=N+1$, future extensions could explore more general team configurations and coordination strategies for $ M\ne N$. 

\section*{Acknowledgment}
This project was partially supported by the National Science Foundation (NSF) through the NRT-AI: Digital Transformation of Development (DToD) program DGE-2125913 and in part by Provably Correct Design of Adaptive Hybrid Neuro-Symbolic Cyber Physical Systems, Defense Advanced Research Projects Agency award number FA8750-23-C-0080. We thank Matt Podolsky and Larry Rohrbough for helpful discussions.

\appendices
\section{Training Parameters and Algorithm}
\label{appendix:training}

We provide a detailed description of the algorithm used in our implementation below. 
\subsection{Policy Optimization}
To train the policies, we used Proximal Policy Optimization (PPO) with Recurrence for POMDPs.

\begin{algorithm}
\caption{PPO with Recurrence (POMDP)} \label{alg:ppo}
\begin{algorithmic}[1]
\footnotesize
\item[] \textbf{Input:} POMDP $\mathcal{M} = \langle \mathcal{S}, \mathcal{A}, T, R, Z, \rho_0 \rangle$\\
\item[] \textbf{Output:} Policy parameters $\theta$ for $\pi_{\theta}$\\
\item[] \textbf{Data:} \texttt{total\_timesteps}, \texttt{frames\_per\_batch}, \texttt{num\_epochs}\\
\STATE Initialize $\theta, \phi$ for $\pi_{\theta}, V_{\phi}$ randomly;
\STATE \texttt{collected\_timesteps} $\leftarrow 0$;
\WHILE{\texttt{collected\_timesteps} $<$ \\ \texttt{total\_timesteps}}
    \STATE $D \leftarrow \{\}$;
    \STATE $\texttt{batch\_collected\_timesteps}\leftarrow 0$;
    \WHILE{\texttt{batch\_collected\_timesteps} $<$ \\ \texttt{frames\_per\_batch}}
        \STATE $\tau \leftarrow \{\}$;
        \STATE Sample $s_0 \sim \rho_0$, $o_0 \sim Z(s_0)$;
        \FOR{$t \leftarrow 0$ to $T - 1$}
            \STATE $h_t \gets [o_0, \ldots, o_t]$;
            \STATE $a_t \sim \pi_\theta(h_t)$;
            \STATE $s_{t+1} \sim T(s_t, a_t)$, $o_{t+1} \sim Z(s_{t+1})$;
            \STATE $r_t \gets R(s_t, a_t, s_{t+1})$;
            \STATE $\tau  \leftarrow \tau \cup (h_t, a_t, r_t)$ to $\tau$;
            \STATE $\texttt{batch\_collected\_timesteps} \leftarrow$  \\ \texttt{batch\_collected\_timesteps} $+ 1$;
        \ENDFOR
        \STATE $D \leftarrow D \cup \tau$;
    \ENDWHILE
    \STATE $\theta' \leftarrow \theta$;
    \FOR{$i = 1$ to \texttt{num\_epochs}}
        \STATE $\delta_t \gets r_t + \gamma V_\phi(h_{t+1}) - V_\phi(h_t)$;
        \STATE $\hat{A}_t \gets \sum_{l=0}^{T - t} (\gamma \lambda)^l \delta_{t+l}$;
        \STATE $\hat{V}_t \gets \hat{A}_t + V_\phi(h_t)$;
        \STATE $e_t \gets \text{entropy}(\pi_\theta(h_t))$;
        \STATE $\xi_t \gets \frac{\pi_{\theta'}(a_t \mid h_t)}{\pi_\theta(a_t \mid h_t)}$;
        \STATE $L^{\text{CLIP}}_t \gets \mathbb{E}_t \left[\min\left(\xi_t \hat{A}_t,\ \text{clip}(\xi_t,\ 1-\epsilon,\ 1+\epsilon) \hat{A}_t\right)\right]$
        \STATE $L^{VF} \gets \mathbb{E}_t \left[ \text{SmoothL1}\left(V_\phi(h_t),\ \hat{V}_t\right) \right]$
        \STATE $L \gets L^{\text{CLIP}}_t-c_0 L_t^{VF}+c_1 \mathbb{E}_t[e_t]$;
    \ENDFOR
    \STATE $\theta \leftarrow \theta'$ ;
\ENDWHILE
\STATE \textbf{return} $\theta$
\end{algorithmic}
\end{algorithm}

In our implementation, we use a Tanh Normal distribution for the sampled action (line 12 in Algorithm~\ref{alg:ppo}), combined with action normalization.

\subsection{Hyperparameters}

Table~\ref{tab:hyperparams_ppo} defines the values we used during training of the rescuer team policy.

\begin{table}[t]
\centering
\caption{Hyperparameters (PPO)}
\label{tab:hyperparams_ppo}
\begin{tabular}{l l}
\hline
\textbf{Hyperparameter}             & \textbf{Value}          \\
\hline
Frames per batch                    & $1024$                   \\
Sub-batch size                      & $256$                    \\
Number of epochs                    & $10$                      \\
Discount factor (\(\gamma\))        & $0.99$                    \\
GAE parameter (\(\lambda\))         & $0.95$                   \\
Clip range (\(\epsilon\))           & $0.2$                     \\
Critic coefficient ($c_0$)          & $0.5$                     \\
Entropy coefficient ($c_1$)         & $0.005$                    \\
1-2 Optimizer                       & Adam                    \\
Learning rate                   & \(3 \times 10^{-4}\)    \\
Betas                           & $(0.9, 0.99)$           \\
Eps                             & $1 \times 10^{-8}$      \\
Weight Decay                    & $0$                      \\
\hline
\end{tabular}
\end{table}


\bibliographystyle{IEEEtran} 
\bibliography{refs} 

\end{document}